\documentclass[aip]{revtex4-1}

\usepackage{graphicx} 
\usepackage{amsmath} 
\usepackage{epstopdf}
\begin{document}


\title{Generation of maximally entangled states and coherent control in quantum dot microlenses}



\author{Samir Bounouar$^{1*}$, Christoph de la Haye$^1$, Max Strau\ss$^1$, Peter Schnauber$^1$, Alexander Thoma$^1$, Manuel Gschrey$^1$, Jan-Hindrik Schulze$^1$, Andr\'{e} Strittmatter$^{1,2}$, Sven Rodt$^1$ and Stephan Reitzenstein$^1$}
\email[Corresponding author: ]{samir.bounouar@tu-berlin.de}
\affiliation{\em Institut f\"ur Festk\"orperphysik, Technische Universit\"at Berlin, 10623 Berlin, Germany}
\affiliation{\em Present address: Abteilung f\"ur Halbleiterepitaxie, Otto-von-Guericke Universit\"at, 39106 Magdeburg, Germany}


\date{\today}

\begin{abstract}
The integration of entangled photon emitters in nanophotonic structures designed for the broadband enhancement of photon extraction is a major challenge for quantum information technologies. We study the potential of quantum dot (QD) microlenses to act as efficient emitters of maximally entangled photons. For this purpose, we perform quantum tomography measurements on InGaAs QDs integrated deterministically into microlenses. Even though the studied QDs show non-zero excitonic fine-structure splitting (FSS), polarization entanglement can be prepared with a fidelity close to unity. The quality of the measured entanglement is only dependent on the temporal resolution of the used single-photon detectors compared to the period of the excitonic phase precession imposed by the FSS. Interestingly, entanglement is kept along the full excitonic wave-packet and is not affected by decoherence. Furthermore, coherent control of the upper biexcitonic state is demonstrated.
\end{abstract}

\pacs{}

\maketitle


Most photonic quantum computation \cite{knill} and quantum communication protocols \cite{wheis} rely on the availability of highly entangled photon pairs. Moreover, entanglement plays the pivotal role in linking the nodes of quantum networks \cite{kimble}. In this context, the excitonic-biexcitonic radiative cascade in quantum dots (QDs) has proved to be a very promising candidate for the generation of polarization entangled photon pairs, in particular because of the triggered emission of photons~\cite{ondem,bounouar}. The main obstacle encountered in the generation of entangled photon pairs with QDs has been the finite excitonic fine-structure splitting (FSS), giving a "which-path" information on the exciton-biexciton radiative cascade \cite{akopian}. In recent years various growth and post-growth techniques have been developed to reduce the FSS. Efforts were done in approaches as diverse as epitaxial growth in (111) direction \cite{versteegh}, growth of highly symmetric GaAs QDs \cite{kapon,Linz,Amand,juska,juska2}, rapid thermal annealing~\cite{young}, external piezo-applied stress \cite{zhang} or electric \cite{benett} and magnetic field tuning \cite{steven,see}. Despite some of them being successful, they are technologically demanding and can negatively affect the quality of the emission, reducing the quantum efficiency \cite{cohsteven} and the spin coherence \cite{benett}. A second issue is the efficient broadband extraction of entangled pairs emitted by a semiconductor QD located in the high refractive index host material. While cavity enhanced emission of entangled photon pairs was achieved using a QD coupled to narrow-band hybridized micropillar cavity modes \cite{dousse}, the scalability and the reproducibility of this very demanding concept is still a non-solved issue. Since the spectral separation between the excitonic and the biexcitonic transitions, namely the biexciton binding energy, is typically in the range of a few  meV, solutions featuring broadband enhancement of photon extraction are most suitable for the realization of QD based entangled photon-pair sources. Over the last decade, a few approaches including photonic wires and microlenses \cite{reimer,claudon,gschrey} tackled this challenge and are good candidates for the production of entangled photon pairs.   

We present here a study on single semiconductor QDs integrated deterministically into microlenses \cite{gschrey}. Since these structures allow for a broadband extraction of the excitonic (X) and biexcitonic (XX) photons as well as for enhanced focusing of the resonant laser\cite{bounouar2}, they are very interesting structures for applications in the field of photonic quantum information technology. We show that two key requirements are fulfilled by these nanostructures. Firstly, by applying pulsed resonant two-photon excitation of the biexciton, we show that the quantum dot upper-state can be coherently addressed and controlled. Secondly, time resolved quantum tomography is performed on photon pairs emitted by the radiative XX-X cascade of the QD. We take advantage of Heisenberg's relation, expressing that higher temporal resolution in determining the dynamics of the XX-X decay implies larger uncertainty in energy which can be larger than the related excitonic fine structure of the QD. In this situation the ''which-path information'' is lost and quantum entanglement of the paired photons can be measured even in the presence of a FSS. The observed degree of entanglement is actually solely limited by the detectors temporal resolution which has to be compared with the inverse precession frequency of the excitonic phase imposed by the FSS. The latter feature is demonstrated by performing quantum tomography on two QD-microlenses with FSS of 16 $\mu$eV and 30 $\mu$eV, respectively. In both cases, photons emitted by the XX-X cascade remain maximally entangled during the radiative decay and are not affected by decoherence. For practical purposes, we provide an estimation of the entanglement degree as a function of the time window applied for the post-selection of the exciton wavepacket.

\begin{figure}[htbp]

\centerline{\includegraphics[width=\linewidth]{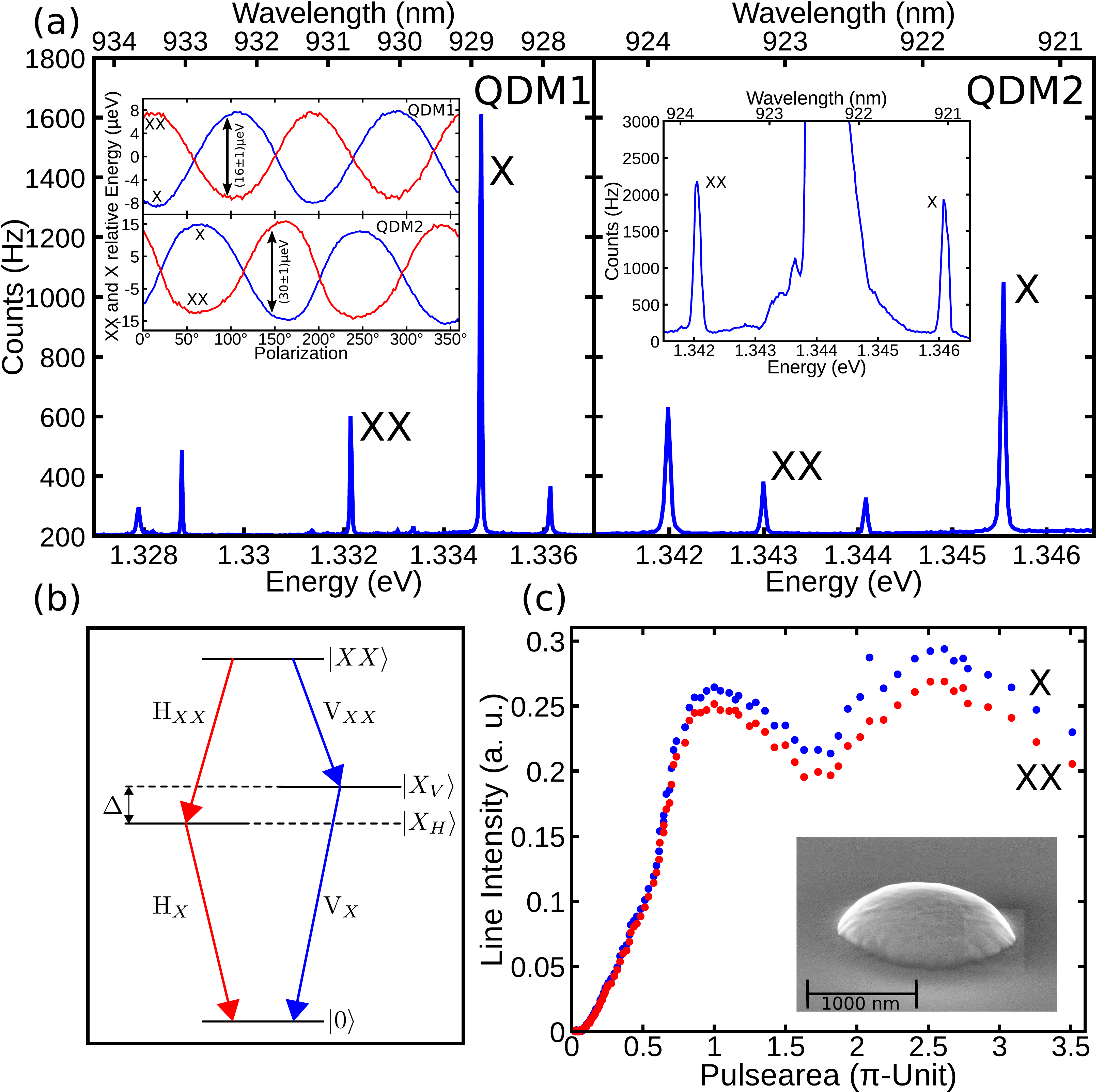}}
\caption{\footnotesize (a) $\mu$PL spectra of QDM1 (left panel) and QDM2 (right panel) under non-resonant excitation (left panel inset: relative energy of the X and XX lines as a function of the detection polarization angle for QDM1 and QDM2. A FSS of 16 $\mu$eV for QDM1 and 30 $\mu$eV for QDM2 are determined by a sinusoidal fit of the experimental data). Right panel inset: exemplary $\mu$PL spectrum of a QD-microlens under resonant two-photon excitation. (b) Scheme of the relevant states in a QD for the generation of entangled photon pairs. (c) $\mu$PL intensity of the XX transition as a function of the two-photon resonant pulse area. (d) SEM image of a deterministically fabricated QD-microlens.}
\label{fig:fig1}

\end{figure}

Our experiments are carried out on self-assembled InGaAs/GaAs QDs grown by metal-organic chemical vapor deposition. The QDs are integrated into microlenses with a backside distributed Bragg reflector by 3D in-situ electron-beam lithography~\cite{gschrey}. A scanning electron microscopy (SEM) image of such a device is displayed in the inset of Fig.~\ref{fig:fig1}~(c). Two different QD-microlenses dubbed QDM1 and QDM2 are studied in the following. Fig.~\ref{fig:fig1}~(a) shows $\mu$PL spectra of QDM1 and QDM2 under non-resonant excitation at 532 nm. Two emission lines in each spectrum are identified as X and XX transitions of the target QD. The inset of Fig.~\ref{fig:fig1}~(a) (right panel inset) shows the typical emission spectrum of a QD-microlense under resonant pulsed two-photon excitation of the biexciton. Fig.~\ref{fig:fig1}~(c) presents the power dependence of the X and XX intensities as a function of the applied pulse area. The laser is placed at an energy resonant to the virtual state enabling the two-photon excitation of the biexciton. This excitation scheme has become a well established and powerful technique \cite{winik,bounouar,jayakumar}, and is nowadays considered as a critical prerequisite for the coherent generation of entangled photons by QDs. We perform these measurements in confocal configuration with a resonant excitation setup. Typical Rabi oscillations of the biexcitonic and excitonic line intensities are observed when the laser pulse area is increased, accounting for the coherent control of the QD biexcitonic state in the Bloch-sphere \cite{stuffler}. The first maximum of the curves plotted in Fig.~\ref{fig:fig1}~(c) represents the first inversion of the biexcitonic population ($\pi$-pulse), and the ideal operation point of the source. The FSS of QDM1 and QDM2 is determined by polarization-dependent $\mu$PL spectroscopy. The inset of Fig.~\ref{fig:fig1}~(a) shows the relative energies of X (blue curve) and  XX (red curve) as a function of the detection angle in linear polarization. Sinusoidal fits to the experimental data yield a FSS of ($15\pm$1)$\mu$eV for QDM1 and $(30\pm$1) $\mu$eV for QDM2, respectively.  

\begin{figure}[htbp]

\centerline{\includegraphics[width=\linewidth]{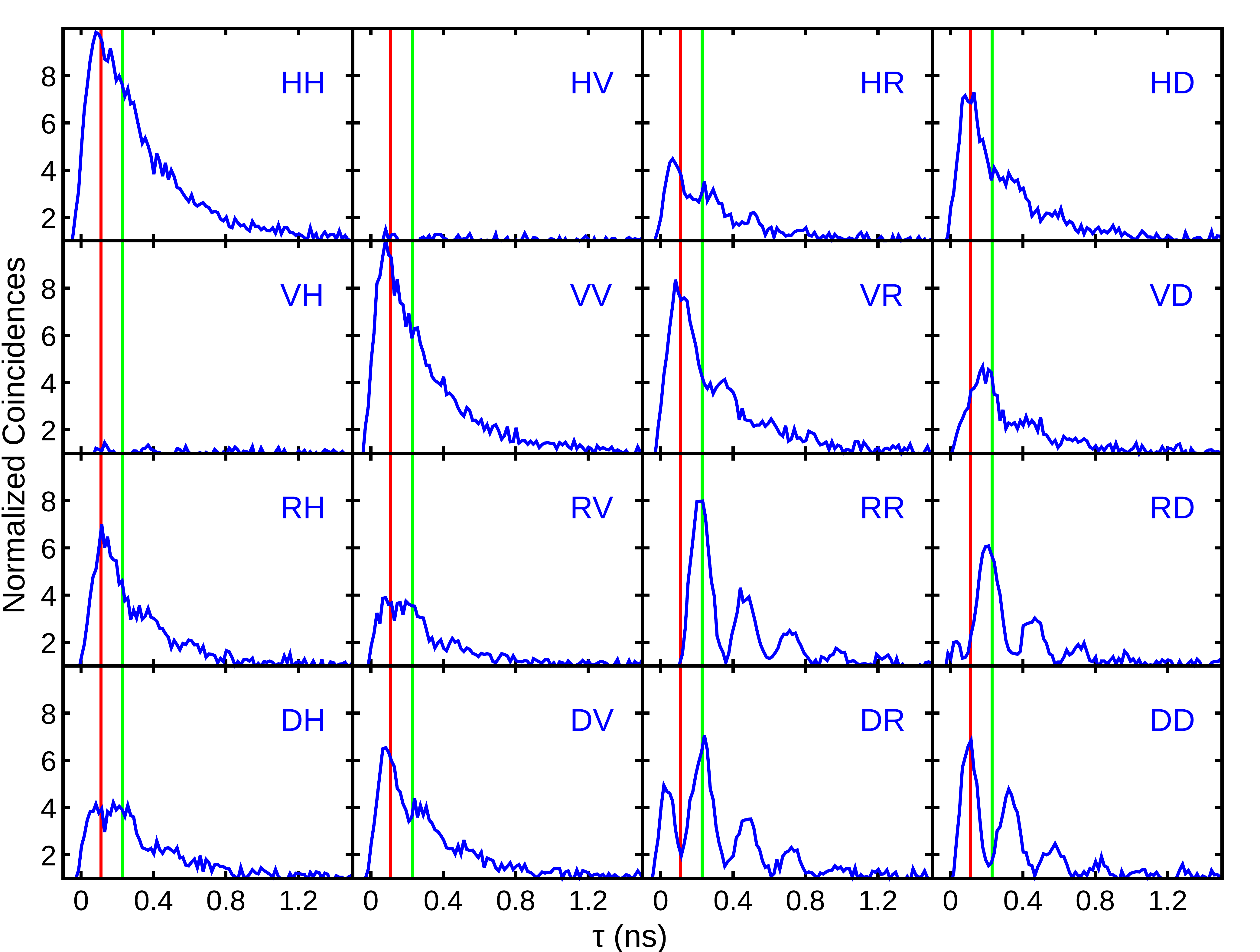}}
\caption{\footnotesize 16 time resolved polarization-dependent correlation measurements used for the quantum tomography for QDM1. The red (green) line represents the time bin used for the density matrix reconstruction noted $\rho_1$ ($\rho_2$) in the following. }
\label{fig:fig2}

\end{figure}

In QDs with a non-zero FSS, the spin up and spin down exciton states (respectively labelled $|X_H\rangle$ and $|X_V\rangle$ in Fig.~\ref{fig:fig1}~(b)) are non-degenerate and are no eigenstates of the system. The exciton state will therefore evolve with time. One can write the resulting two-photon state as follows \cite{cohsteven}:

\begin{align*}
\label{eq:solve}
|\psi (\tau)\rangle=\frac{1}{\sqrt{2}}(|HH\rangle+e^{i\Delta \tau}|VV\rangle),
\end{align*}

with $\Delta$ the frequency corresponding to the FSS energy and $\tau$ the time delay between the excitonic and biexcitonic photons. This state can be rewritten in the diagonal basis (D/A) to show that it oscillates between $|\phi^+\rangle=\frac{1}{\sqrt{2}}(|DD\rangle+|AA\rangle)$ and $|\phi^-\rangle=\frac{1}{\sqrt{2}}(|DA\rangle+|AD\rangle)$. One obtains a similar result in the circular basis (R/L). This means that the excitonic phase evolution, which leads to this oscillation, can be tracked by correlating the photons when they are projected in these two bases.
 
For the quantum tomography measurements we use a time resolved polarization-dependent cross-correlation setup. Photons from the coherently driven (under $\pi$-pulse condition) XX-X cascade are polarization-projected in the 3 complementary bases (H/V, D/A and R/L), and the coincidence rate is measured as a function of the delay between the XX photons and the X photons. The 16 measurements necessary for the full quantum tomography and the corresponding normalized correlation functions are plotted in Fig.~\ref{fig:fig2}. The observed time dependent oscillations due to the excitonic phase evolution are discussed above and occur when both photons are projected in the right circular polarization R or in the diagonal polarization D. On the other hand, the measurements in the linear basis (HH and HV), showing no oscillation, exhibit classical correlations. The temporal resolution of the setup is estimated to be 100 ps (full width at half maximum) and each coincidence time bin is 4 ps.

\begin{figure}[htbp]
\centerline{\includegraphics[width=\linewidth]{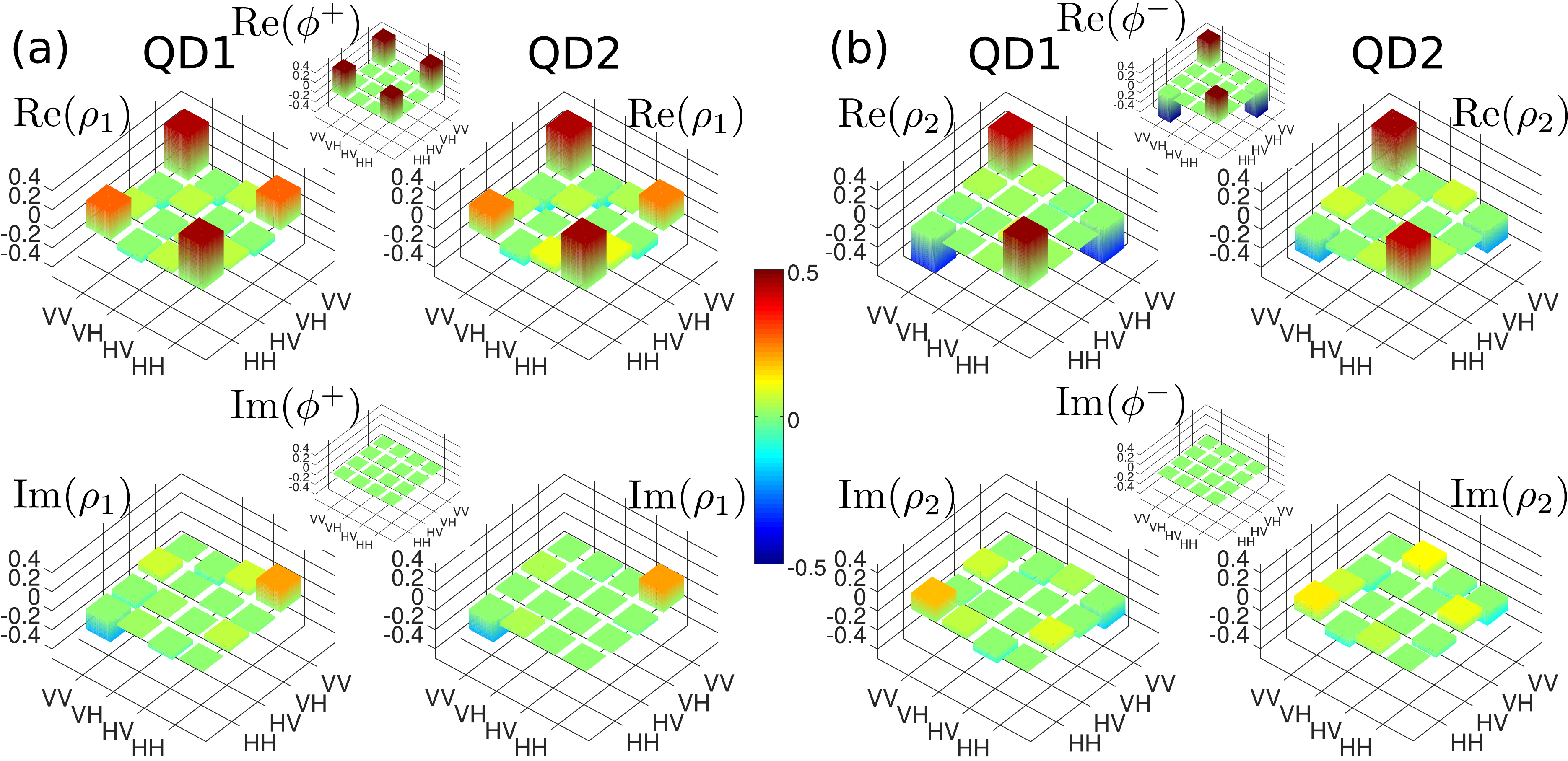}}
\caption{\footnotesize (a) Density matrices reconstructed for a delay corresponding to the first maximum of the DD coincidence curve (noted $\rho_1$, 4 ps selection time window), for QDM1 (left panel) and for QDM2 (right panel). Real parts are displayed on the top and imaginary parts on the bottom part of the graph. Theoretical real parts and imaginary parts of the maximally entangled state $|\phi^+\rangle$ are plotted between the matrices of QDM1 and QDM2, as a reference. (b) Same density matrices reconstructed for a delay corresponding to the first minimum of the DD coincidence curve (noted $\rho_2$, 4 ps selection time window). Theoretical real parts and imaginary parts of the maximally entangled state $|\phi^-\rangle$ are plotted between the matrices of QDM1 and QDM2, as a reference.}
\label{fig:fig3}

\end{figure}

Based on the experimental data presented in Fig.~\ref{fig:fig2}, the density matrices of the generated two-photon states are reconstructed using a maximum-likelihood estimation. Fig.~\ref{fig:fig3} shows the reconstructed density matrices obtained for the first maximum (Fig.~\ref{fig:fig2}, red line)  and the first minimum (Fig.~\ref{fig:fig2}, green line) observed on the DD curve of Fig.~\ref{fig:fig2} (red line marked). Fig.~\ref{fig:fig3}(a) corresponds to the state of the QD directly after the emission of the biexcitonic photon (left panel for QDM1 and right panel for QDM2). For the sake of comparison, the ideal density matrix (real part and imaginary part) of $|\phi^+\rangle$ between the experimental density matrices for QDM1 and QDM2 is displayed in Fig.~\ref{fig:fig3}~(a). The fidelity ($F(\phi^+)=Tr(\sqrt{\sqrt{\rho_1}.\rho(\phi^+).\sqrt{\rho_1}})^2$) of the experimental density matrix ${\rho_1}$ to $|\phi^+\rangle$ is estimated as 0.73$\pm$0.03 for QDM1 and 0.69$\pm$0.04  for QDM2. Since the phase in QDM1 is evolving slower than for QDM2, the setup is able to better resolve the oscillation for QDM1 which also shows a higher degree of entanglement. At longer delays, the QD state rotates towards $|\phi^-\rangle$. Fig.~\ref{fig:fig3}~(b) shows the reconstructed density matrices obtained for the first minimum of the DD coincidence curve (green line marked on Fig.~\ref{fig:fig2}). They resemble the $|\phi^-\rangle$ state (represented in inset of Fig.~\ref{fig:fig3}~b)). $F(\phi^-)$ is estimated to 0.80$\pm$0.03 for QDM1 and 0.68$\pm$0.04 for QDM2.

\begin{figure}[htbp]

\centerline{\includegraphics[width=\linewidth]{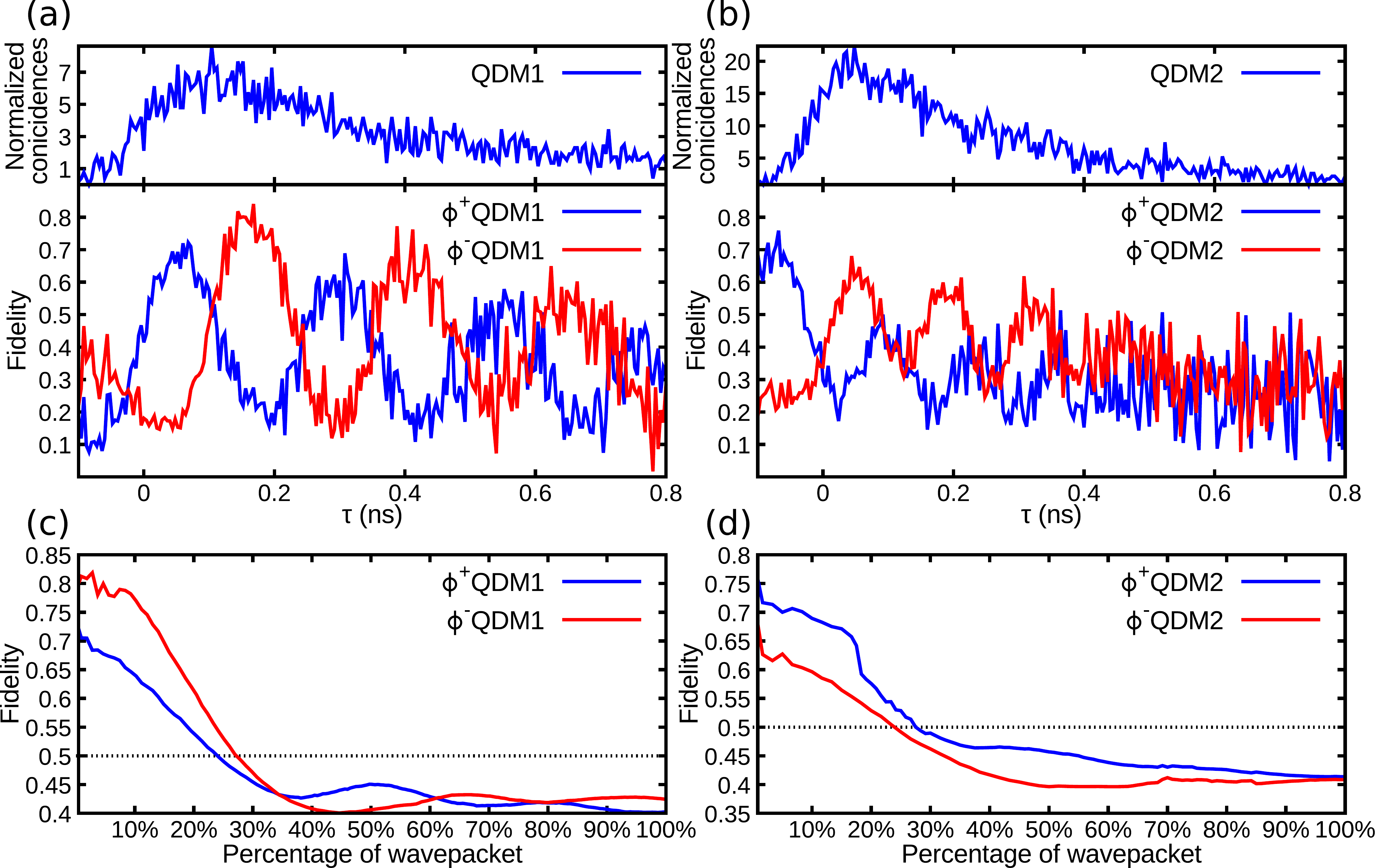}}
\caption{\footnotesize (a) Data obtained for QDM1: (upper panel) Normalized coincidences between a biexcitonic photon and a triggering excitonic photon. (lower panel) Measured fidelities to the maximally entangled Bell states (blue curve for $|\phi^+\rangle$, and red curve for $|\phi^-\rangle$). (b) Same data obtained with QDM2. (c) Measured fidelity as a function of the percentage of the post-selected excitonic wavepacket for QDM1, (d) for QDM2.}
\label{fig:fig4}

\end{figure}

Fig.~\ref{fig:fig4}~(a) depicts the time evolution of the fidelity to the two Bell states, $F(\phi^+)$ and $F(\phi^-)$ of the two-photon state. As expected, because of the excitonic phase evolution, the entangled two-photon state evolves between $|\phi^+\rangle$ and $|\phi^-\rangle$. Interestingly, the oscillations can be clearly observed along the whole exciton wavepacket (plotted on the top panel of Fig.~\ref{fig:fig4}~(a)), indicating that the entanglement of the QD state is mostly unaffected by decoherence. Fig.~\ref{fig:fig4}~(c) shows $F(\phi^+)$ and $F(\phi^-)$ as a function of the time windows used for the tomography. The fidelities for both quantum dots are decreasing quickly as the time window is enlarged and for a post-selection exceeding 20\% of the total exciton wavepacket, no entanglement can be observed any more. An optimal working point is obtained by post-selection of around 15\% of the excitonic photons, presenting here a good compromise between fidelity to the Bell state and photon count rate.

\begin{figure}[htbp]

\centerline{\includegraphics[width=\linewidth]{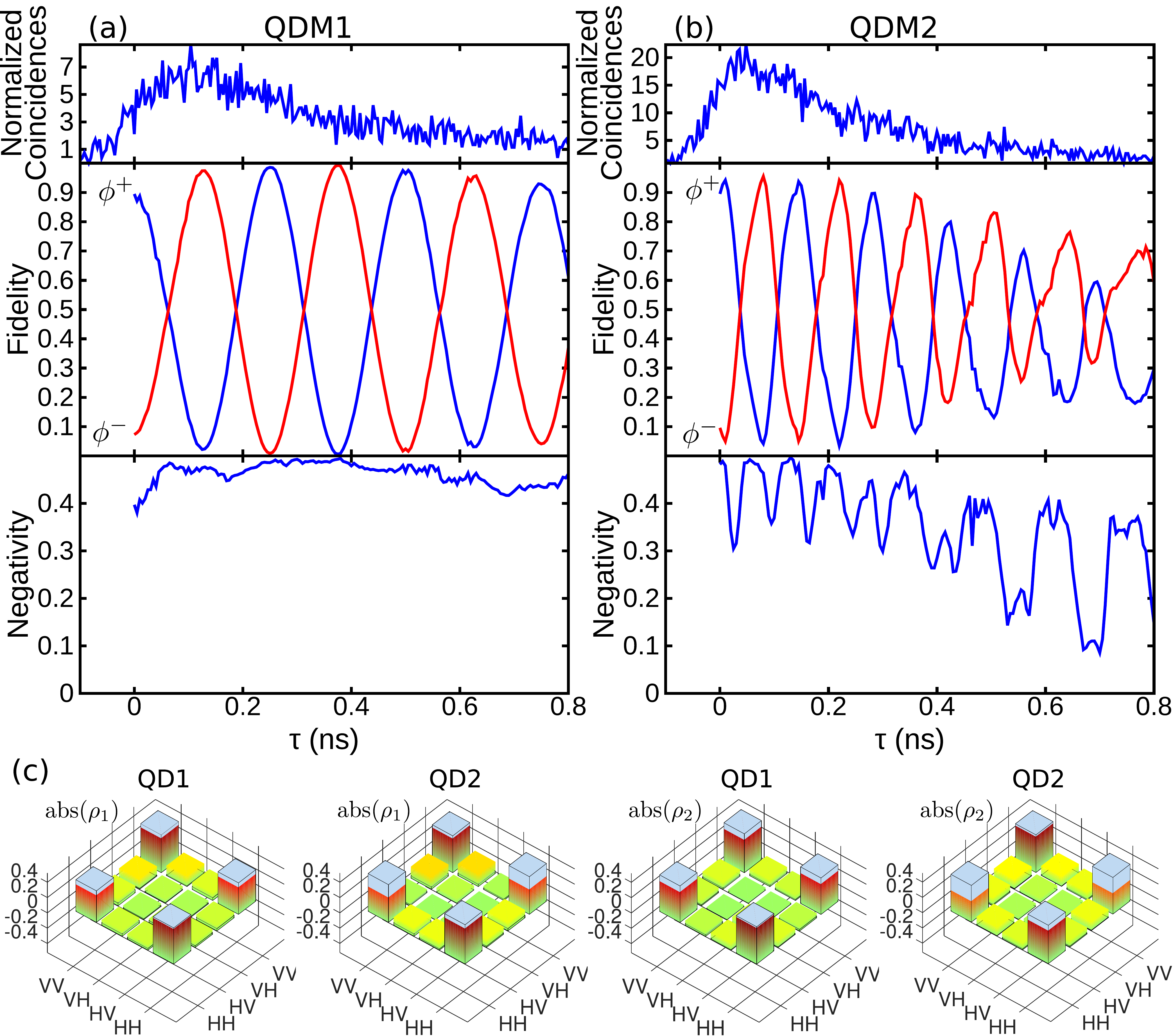}}
\caption{\footnotesize (a) Deconvoluted data obtained for QDM1. Upper panel: normalized coincidences between an excitonic photon and a triggering biexcitonic photon. Intermediate panel: Measured fidelities to the maximally entangled Bell states (blue curve for $|\phi^+\rangle$, and red curve for $|\phi^-\rangle$). Lower panel: Negativity of the two-photon state as a function of the delay after deconvolution, (b) Deconvoluted data obtained with QDM2, (c) Superposed norms of the density matrices reconstructed before convolution (in color), and corrected after deconvolution (in blue) for QDM1 and QDM2.}
\label{fig:fig5}

\end{figure}

In order to evaluate the actual quality of the entanglement between the photons emitted by the QD, we deconvoluted the data from the time response of the experimental tomography setup with a temporal resolution of 100~ps. The theoretical polarization-dependant cross-correlation function~\cite{winik} is convoluted to a Gaussian curve (100 ps full width at half maximum), as measured from the setup response. The resulting curve is fitted to the 16 tomography data curves. All the function parameters, such as the polarization angles, the FSS frequency and the decay time, were obtained from experiment and are kept constant. Only a multiplying factor and an offset are left as free parameters for the data fitting (see supplementary material). The deconvoluted fit function is used in order to reconstruct the new density matrices representing the emitted two-photon states without the effect of the experimental resolution. Fig~\ref{fig:fig5} shows the results provided by the quantum tomography after deconvolution. Fig.~\ref{fig:fig5}~(a) shows the decay curve (upper panel), the fidelity to the Bell states (middle panel) and the negativity (lower panel) for QDM1. The fidelities are showing oscillations with an amplitude very close to unity without damping along the full wavepacket, showing that QDM1 emits nearly perfectly entangled photons. This is confirmed by the negativity (lower panel), quantifying the separability of the density matrix. A value close to 0.5 (maximal entanglement) is found up to a delay of 0.8 ns.

Fig.~\ref{fig:fig5}~(b) shows very similar results for QDM2 with larger FSS. For instance, oscillations with close to unity amplitude can also be observed on the fidelity curves. For this QD more pronounced damping of the oscillations is observed indicating that the quality of the entanglement is significantly reduced for delays larger than 0.5 ns. The decrease of the negativity with respect to the delay is consistent with this observation. Even if simultaneous jumps of the excitonic and biexcitonic phases do not affect the QD entanglement, cross-dephasing processes, such as exciton spin 'flip-flop' processes, could be the reason for such a degradation of the entanglement quality for this particular QD. Moreover, the deconvolution is not fully successful at suppressing the periodic drops of the negativity. This can be attributed to the frequency of the phase rotation being too close to the experimental resolution.

In conclusion, we have shown that determinstically fabricated QD-microlenses with broadband photon extraction are very suitable for the reliable generation of entangled photon pairs. This is demonstrated by two-photon excitation of the biexciton in QD-microleneses where for finite FSS the entanglement fidelty is only limited by the experimental time resolution. Interestingly, the decoherence of XX and X is not affecting their fidelity to the Bell states. These achievements open the possibility of using QDs showing FSS exceeding 10 $\mu$eV in photonic quantum technology schemes, but at the price of an event ''post-selection''. In this respect, the microlenses or other high efficiency broadband nanophotonic elements are of great interest. However, a reduction of the FSS is still of great importance since it allows for the use of less narrow post-selection windows, slower detectors and shorter integration times. Entangled photon pair emission from such optical nanodevices represent a significant step towards the practical and generalized realization of entanglement swapping or teleportation experiments which are key requirements for long-distance quantum communication and photonic quantum computation.

The research leading to these results has received funding from the German Research Foundation via Projects No. RE2974/4-1, 
No. RE2974/12-1, SFB 787, and from the German Federal Ministry of Education and Research (BMBF) through the VIP-project 
QSOURCE (Grant No. 03V0630).

\end{document}